# Optical phonon modes assisted thermal conductivity in p-type ZrIrSb Half-Heusler alloy: A combined experimental and computational study

Kavita Yadav[1], Saurabh Singh[2], Tsunehiro Takeuchi[2], and K. Mukherjee[1]

[1]School of Basic Sciences, Indian Institute of Technology, Mandi, Himachal Pradesh-175005, India

[2]Research Centre for Smart Energy Technology, Toyota Technological Institute, Nagoya, 468-8511, Japan

**Abstract**

Half Heusler (HH) alloys with 18 valence electron count have attracted significant interest in the area of research related to thermoelectrics. Understanding the novel transport properties exhibited by these systems with semiconducting ground state is an important focus area in this field. Large thermal conductivity shown by most of the HH alloy possesses a major hurdle in improving the figure of merit (ZT). Additionally, understanding the mechanism of thermal conduction in heavy constituents HH alloys is an interesting aspect. Here, we have investigated the high temperature thermoelectric properties of ZrIrSb through experimental studies, phonon dispersion and electronic band structure calculations. ZrIrSb is found to exhibit substantially lower magnitude of resistivity and Seebeck coefficient near room temperature, owing to existence of anti-site disorder between Ir/Sb and vacant sites. Interestingly, in ZrIrSb, lattice thermal conductivity is governed by coupling between the acoustic and low frequency optical phonon modes, which originates due to heavier Ir/Sb atoms. This coupling leads to an enhancement in the Umklapp processes due to the optical phonon excitations near zone boundary, resulting in a lower magnitude of $\kappa_L$. Our studies point to the fact that the simultaneous existence of two heavy mass elements within a simple unit cell can substantially decrease the lattice degrees of freedom.

## 1. Introduction

Thermoelectric (TE) materials are important as they are used to extract waste heat and convert it into useful electrical energy. This unique property of thermoelectric materials has attracted immense attention among researchers in recent years from the viewpoint of interesting electronic band structure and due to their potential technological applications in home heating appliances, temperature sensors, thermoelectric cooler for electronic devices, cooling infrared sensors [1-5]. In this aspect, semiconducting half-Heusler (HH) alloys have emerged as an important class of materials [6-8]. These materials are also free of toxic elements and have high thermal stability [9, 10]. The HH alloys crystallize in cubic structure (space group *F-43m*) and typically have the formula XYZ. The elements X, Y, and Z occupy the Wyckoff position 4*a* (0, 0,0), 4*c* (0.25, 0.25, 0.25), and 4*b* (0.5, 0.5, 0.5), respectively, and leaving the position 4*d* (0.75, 0.75, 0.75) as vacant. Each occupied site can be tuned independently, through substitutions, in order to optimize the TE system. These alloys are categorized into different groups, depending upon the valence electron count (VEC) [6, 11]. For example, LiMgZ (Z = N, P) have 8 VEC in the primitive unit cell. These alloys are non-magnetic and semiconducting [12]. Based upon the theoretical calculations, Kandpal *et al.* have formulated a rule that the HHs with VEC = 8 or 18 exhibit semiconducting behaviour [11]. Further, depending on the magnitude of band gaps, some of them display exceptional thermoelectric properties such as large Seebeck coefficient ($S$) and reasonably small electrical resistivity ($\rho$) at high temperatures [13-15]. But the presence of high thermal conductivity ($\kappa_{total}$) in these systems, limits their capability as a good thermoelectric material. Generally, the transport of heat is mainly through the lattice degrees of freedom i.e., there is large contribution from lattice thermal conductivity ($\kappa_L$) [6-8, 13-15]. $\kappa_L$ can be effectively reduced by inducing mass fluctuations or strain field effects. These effects can be introduced into the HH systems by substituting or doping heavy elements like Sb, Hf, Zr etc. For instance, in XNiSn-based HH systems, (where X = Ti, Hf, Zr), substitution of multiple elements with different masses, at the X as well as Ni sites, has led to significant reduction in the magnitude of $\kappa_L$ [16]. On the other hand, similar effect is observed in both ZrCoSb and TiCoSb, where $\kappa_L$ is dramatically decreased with increasing Sn and Fe content at Sb and Co site, respectively [17, 18]. Hence, it can be said that substitution of different or heavier mass elements can lead to a substantial decrement in $\kappa_L$ in HH alloys.

In order to unravel the physics behind the mechanism of thermal conduction, a systematic study of phonon properties of the thermoelectric material is crucial [19-23]. The phonon calculations are useful in extracting the information about the stability of the crystal and contributions from the different phonon modes to $\kappa_L$. Hence, computational study of phonon dispersion along with the phonon density of states (P-DOS) is very important for the complete description of the heat transport mechanism in a thermoelectric material. In this regard, there has been several studies on semi-conducting HH alloys, focussing on their phonon dispersion and P-DOS. Most of these investigations have been carried out using first principles calculations. From these calculations, it has been determined that the fundamental factors significantly contributing to a lower $\kappa_L$ are weak interatomic interactions, heavy element substitution and large Gruneisen parameters [24-26]. It has also been noted that there is dominant contribution from the acoustic phonon modes due to their relatively higher group velocity in comparison with the optical phonon modes. Furthermore, in some cases it has been observed that there is a strong coupling between the low frequency optical and acoustic modes, leading to ultralow $\kappa_L$. Large unit cell, blocked crystal structure, heavy mass elements can be the possible reasons for latter observation [24-26]. However, simple crystal structure of HH alloys possesses as a major challenge in this. Hence, it is vital to understand the individual contributions from acoustic and optical phonon modes to $\kappa_L$, in a simple unit cell. This, in turn, will be helpful in exploring and designing new low $\kappa_L$ based HH alloys. In view of the above, ZrIrSb HH alloy provides a fertile ground for investigation due to its heavy constituent elements and simple crystal structure. Theoretically, this alloy is predicted to have semiconducting ground state and interesting conduction band structure, which can give a significant thermopower at high temperature [27]. But to the best of our knowledge, this alloy has not been explored experimentally.

Hence, in this manuscript, we have studied the high temperature thermoelectric properties of ZrIrSb in the temperature regime of 300-950 K using both experimental and computational tools. Inspite of a wide band gap in ZrIrSb, our experimental results indicate a lower magnitude of the $\rho$ at room temperature. This anomalous observation can be attributed to the presence of anti-site disorder between Ir/Sb and vacant sites. The observed positive $S$ is well understood in terms of effective mass of the charge carriers. In this alloy, the heat conduction is via coupling between the low frequency optical and acoustic phonon modes, arising due to the heavier elements Ir/Sb. This coupling results in a strong phonon-phonon scattering due to the presence of significant Umklapp processes. It leads to lower magnitude of $\kappa_L$ and improvement in ZT in comparison to its analogous alloys.

## 2. Experimental and computational details

Polycrystalline alloy ZrIrSb is synthesized by arc melting (in argon atmosphere) the stoichiometric ratio of high purity constituent elements (>99.9%) procured from Sigma Aldrich (Zr, Sb) and Alfa Aesar (Ir). The ingot is re-melted several times to ensure the homogeneity of the alloy. Due to high vapour pressure of Sb, an additional amount of 5% Sb is added to compensate the Sb loss during arc melting. The weight loss after the final melting is < 2%. The obtained ingot is crushed into powder using mortar and pestle. Further, obtained powdered sample is pelletized and sintered using spark plasma sintering technique (SPS) at 1473 K for 5 Min under 50 MPa in argon atmosphere. The phase purity and crystal structure is confirmed through room temperature X-ray diffraction (XRD) performed using Bruker D8 Advance (Cu-Kα source). The compositional analysis is carried out by using the electron probe micro analysis (EPMA) (make: JEOL JXL-8230). $S$ and $\rho$ are measured by the steady state method and four probe measurement technique, respectively using the home-made setups [28]. $\kappa_{total}$ measurement is done by using laser flash analysis (NETZSCH LFA 457). Archimedes principle is used to determine the experimental density of the alloy. Longitudinal and shear sound velocities at room temperature are estimated by using an ultrasonic velocity measurement system (ultrasonic transducer, Olympus).

To understand the charge transport behaviour, the electronic structure calculations of ZrIrSb is performed by using the first principles density functional theory (DFT) calculations. The generalized gradient approximation (GGA) of Perdue–Burke–Ernzerhof (PBE) is used as exchange correlation functional, as implemented in WIEN2K code [29, 30]. For the convergence of the electronic total energy, a criterion of $10^{-4}$ Ry/cell is selected. For self-consistency field calculations and electronic density of states (DOS), a dense *k*-point mesh of 15 X 15 X 15, and 20 X 20 X 20 points, respectively is considered [31]. The electronic band dispersion plot is obtained along the high symmetric *k*-points corresponding to the space group *F-43m* (216). The experimental lattice parameter is used to perform the DFT calculations. This parameter is obtained after the Rietveld refinement of the room temperature XRD pattern of ZrIrSb. The theoretical lattice parameter from both VASP and WEIN2K code along with Wyckoff positions of the Zr, Ir and Sb, used for calculations, are given in table 1.

Phonon calculations are performed within the framework of the force constants method implemented in the PHONOPY code [32]. A unit cell and 2 X 2 X 2 supercell are used for the force calculations. Real-space force constants of the supercell are determined in the density functional perturbation theory (DFPT) as implemented in the VASP code. The GGA of PBE is employed for the exchange correlation potential. A plane-wave energy cut-off of 500 eV is used for the calculations. The P-DOS and phonon dispersion are estimated from the force constant. For the P-DOS, a 16 × 16 × 16 mesh is used, whereas phonon dispersion is plotted along the high symmetric directions. The calculation of $\kappa_L$ is executed in PHONO3PY code [33] which is based on the first principles linearized Boltzmann transport equation (LBTE) in the single mode relaxation time (SMRT) approximation with same supercell and 11 X 11 X 11 mesh.

## 3. Results and Discussions

### 3.1 Structural characterizations

The room temperature XRD pattern and crystal structure for ZrIrSb is shown in Fig. 1 (a) and (b). To confirm the structural phase, XRD data of the alloy is analysed by using the Rietveld refinement method implemented in Full Prof software and crystal structure is obtained from VESTA software [34, 35]. The obtained structural parameters are listed in table 1. The alloy is found to be in single-phase; with presence of a two minor unidentified impurity peaks (< 2%; marked by asterisk). The Miller indices of the XRD peaks corresponding to the Bragg reflection positions (shown in Fig. 1) are associated with the cubic crystal structure phase (space group: *F-43m*). The obtained lattice parameter (*a*) is in good agreement with the lattice parameter used in the theoretical calculations and are given in table 1. In these types of alloys, from the intensity of (111), (200) and (220) peaks, one can determine the type of disorder in the unit cell [36]. In our case, (111) peak is clearly visible, however, the intensity of (200) peak is found to be negligible as compared to the most intense peak (220). This signifies that there is a partial presence of Ir/Sb at vacant sites, which leads to anti-site disorder between the two sites. The density of ZrIrSb obtained from Archimedes principle is ~ 10.473g/cc. This value is 97% greater than the theortical value. From EPMA analysis, the average composition is found to be $Zr_{1.12\pm0.02}Ir_{0.98\pm0.03}Sb_{0.90\pm0.01}$, which indicates the non-stoichiometric composition of the ZrIrSb alloy.

### 3.2 Electronic band structure of ZrIrSb

In order to know about the ground state of Heusler alloys, self-consistent field calculations are done. The band structures obtained from these calculations are helpful to understand the role of various elements in the formation of valence band (VB) and conduction bands (CB).

The band structure of ZrIrSb along the directions W-L-Γ-X-W-X of the first Brillouin zone (BZ) is presented in Fig. 2 (a). It can be inferred that, valence band maxima (VBM) and conduction band minima (CBM) lie at two different *k*-points i.e., Γ and X point respectively. This indicates the indirect nature of the band gap, with $E_g \sim 1.41$ eV. Here, it can be noted that Γ is triply degenerate and X point is non-degenerate. Similar type of indirect band gaps has been reported in other well-known semiconducting HH systems, such as HfIrSb, ZrNiSn and HfNiSn [19, 27]. Mal-Soon Lee *et al.* had also investigated the electronic band structure of ZrIrSb and the reported similar value of the band gap [27]. From the figure it can be seen that the bands, numbered from 1-4, contribute to transport properties. As reported in Ref. 27, below Fermi level ($E_F$), bands 1 and 2 are formed due to the hybridization between the *d*-orbitals of Zr and Ir, whereas band 3 is formed from the *d*-orbitals of Ir. The most important features of band structure which affect the transport properties i.e., *S* and *ρ* are (a) the degeneracy of VBM and CBM; (b) effective mass ($m^*$) of the charge carriers at the VBM and CBM and (c) variation of DOS in the vicinity of band gap. As noted earlier, VBM at Γ point is triply degenerate. This signifies that more states are available for the occupation of charge carriers near the $E_F$ at Γ point. Above $E_F$, band 4 is also formed due to hybridization between the *d* states of Ir and neighbouring Zr atoms but has a strong Zr *d* character near the X point [27]. However, this band is non-degenerate implying contribution from only one electron pocket at X point. This can affect the overall magnitude of *S*. Figure 2 (b) shows the total DOS (TDOS) and partial DOS (PDOS) plotted as a function of energy. From the figure it can be said that Zr and Ir are the main contributors to DOS around the Fermi level (Fig. 2 (b)). We have also noted a slow variation of DOS with energy. This implies that this alloy should have a lower magnitude of *S*, which has been confirmed experimentally (discussed in detail in section 3.3.1).

### 3.3 Transport properties

### 3.3.1 Seebeck coefficient

Figure 3 (a) represents the temperature dependent *S* in the temperature regime of 300-950 K. At 300 K, the value of *S* is found to be ~ 34.4 µV/K. The positive sign indicates the dominance of holes in the charge transport mechanism. With the increment in temperature, a linear increment in magnitude of *S* is noted till 500 K (denoted by solid red line in Fig. 3 (a)). Above 500 K, *S* increases non-linearly with temperature, reaching maximum value ($S_{max}$) ~ 75.5 µV/K at 950 K. The observed $S_{max}$ is smaller as compared to XCoSb (X= Ti, Zr and Hf) series of alloys [17, 37-38]. Under free electron theory approximation, $S$ and $m^*$ of the charge carriers are directly related to each other and are given by the expression [39, 40]:

$$S = \frac{8\pi^2 k_B^2}{3eh^2} m^* T \left(\frac{\pi}{3n}\right)^{2/3} \ldots\ldots\ldots (1)$$

where $k_B$ is the Boltzmann constant, $e$ is the electronic charge and $n$ is the charge carrier density. As $T$ and $n$ are positive quantities it implies that the sign of $S$ will be decided by the sign of $m^*$. Generally, $m^*$ depends on the band structure of the materials. Additionally, in case of semiconductors both type of charge carriers (electrons and holes) contributes to the $S$. So, the sign of $S$ will be decided by the charge carriers having larger $m^*$. Thus, $m^*$ estimated from Fig. 2 (a) will be useful to make a qualitative understanding about the sign of experimentally obtained value of $S$. Hence, the VBM and CBM near Γ and X $k$-symmetric points are analysed. The curvature of energy curve at a $k$-point decides the value of $m^*$ and is given by [40]

$$m^* = \frac{\hbar}{\left[\frac{d^2 E}{dk^2}\right]}\ldots\ldots\ldots (2)$$

From the above equation, it can be inferred that at a given $k$ point; the $m^*$ of a charge carrier, for a broader energy curve is greater than the narrower energy curve. In order to see the $m^*$ contributions of holes and electrons along different high symmetric directions, $m^*$ has been estimated at Γ point for holes (along K, W, X and L directions) and at X point for electrons (along Γ, W, K and L). The calculated $m^*$ is given in table 2. In the table, K-Γ-K represents the $m^*$ calculated at Γ point along the Γ-K direction. Similar notations are used for other directions as well. It is noted that band 1 is lighter with lesser $m^*$, resulting in higher electrical conductivity of charge carriers in this band. However, band 2 and 3 being degenerate along Γ-X and Γ-L directions have same $m^*$ values. The estimated value of $m^*$ for holes is 0.6 $m_e$ (of band 2 and 3) and 0.21 $m_e$ (band 1) along the X direction. Similarly, for electrons the maximum $m^* \sim 0.6$ $m_e$ is found along Γ direction, indicating higher mobility of electrons in the vicinity of CBM. This indicates that both holes and electrons have similar $m^*$ along X and Γ directions, respectively. However, due to degeneracy of bands 1, 2 and 3 in the vicinity of Γ point, there will be larger contribution from the $m^*$ of holes in $S$. It is in analogy with our experimental results. In comparison to TiCoSb, ZrCoSb and HfCoSb a lower $S$ is observed in this alloy. This is due to the fact that there is contribution from only one Γ-pocket in the latter case, whereas, for the other alloys four bands (at Γ point) contribute to the $S$.

Additionally, the presence of both anti-site disorder between Ir/Sb atoms and vacant sites may contribute to the lower value of *S* in ZrIrSb.

**3.3.2 Temperature response of resistivity**

The temperature response of $\rho$ of ZrIrSb in the temperature regime 300-950 K is shown in Fig. 3 (b). At 300 K, the value of $\rho$ is noted to be ~ 16 mΩ-cm. This value is quite small as compared to the values noted for other XCoSb (X = Zr, Hf and Ti) HH alloys [17, 37- 38]. But it is greater than that reported for other well-known thermoelectric materials (as given in table 3) [41-43]. It is observed that $\rho$ decreases non-linearly as a function of temperature upto 370 K. Above 370 K, a linear behaviour is noted. With further increment in temperature (above 750 K), an increasing trend in $\rho$ value is observed (as shown in inset of Fig. 3(a)). The changes in slope of resistivity curve indicate that there is an existence of different conduction mechanisms in the measured temperature regime. Additionally, as discussed in section 3.3.1, an increment in *S* is noted as the temperature is increased. These observations support the fact that in ZrIrSb, the electrical transport is possibly via delocalization of localized electrons. Therefore, to identify the mode of electronic transport, the obtained data is analysed with different types of models available in literature. One of the well-known modes of conduction is nearest neighbour hopping (NHH) [44]. In this mechanism, there is an existence of constant activation energy ($E_a$), which is responsible for the hopping of charge carrier to the nearest neighbour sites from their initial position. The $\rho$ is expressed as [44]

$$\rho(T) = \rho_0\, e^{\frac{E_a}{k_B T}} \quad\ldots\ldots\ldots\ldots\ldots\ldots (3)$$

where $\rho_0$ is the pre-exponential factor, and *T* is the temperature. In the temperature regime of 370-640 K, the curve is fitted well with equation (3) (shown in upper inset of Fig. 3 (b)). The value of $E_a$ is estimated to be ~ 0.1 eV. However, below 370 K, as mentioned before, a non-linear behaviour is observed. This suggests to a presence of more than one $E_a$ and points toward a different conduction mechanism. Generally, non-linear temperature variation of $\rho$ is expressed as

$$\rho = \rho_0\, e^{\left(\frac{T_0}{T}\right)^{\gamma}} \quad\ldots\ldots\ldots\ldots\ldots\ldots\ldots (4)$$

where $\gamma$ can take value 1/4, 1/2, and 1/3 depending upon the different conduction models and $T_0$ is the characteristic Mott's temperature. The curve in the regime 300-370 K is best fitted (using eqn. 4) for $\gamma=1/4$ (shown in lower inset of Fig. 3(b)). This indicate that conduction in this regime is through Mott's variable range hopping (MVRH), where hopping length decreases with increment in temperature. Also, above 750 K, $\rho$ displays metallic type of conduction (as shown in inset of Fig. 3 (a)). Hence from our analysis, it can be concluded that there is transformation from MVRH to NHH around 370 K, followed by metallic type of conductivity above 750 K.

Additionally, the lower magnitude of $\rho$ (near 300 K) in comparison to ZrCoSb (given in table 3) can be attributed to the presence of anti-site disorder between Ir/Sb atoms and vacant sites in ZrIrSb. Due to this disorder, there is always a probability of finding Ir/Sb atoms at 4d position (vacant sites). This can alter the hybridization between Co and Ir atoms, which is responsible for the formation of band gap. It can either lead to reduction in the band gap or lead to formation of in-gap electronic states near VB and CB. Similar effect of anti-site disorder on $\rho$ behaviour is also noted in ZrNiSn [19, 45]. The presence of this disorder is likely to reconcile the lower magnitude of $\rho$ of ZrIrSb alloy in comparison with ZrCoSb. This, along with the non-stoichiometry of the alloy (as reported in section 3.1), can result in the formation of impurity bands near conduction band, possibly leading to the observed hopping type of conduction.

### 3.3.3. Thermal conductivity

The temperature response of $\kappa_{total}$ is shown in Fig. 4 (a). Near 300 K, the magnitude of $\kappa_{total}$ is ~ 9.3 W/mK. In comparison with XCoSb family (as listed in table 4), a significant reduction in room temperature $\kappa_{total}$ is noted. From Fig. 4 (a), it can be inferred that $\kappa_{total}$ decreases with temperature till 500 K. However, a slight increment in $\kappa_{total}$, followed by hump is noted in 500-650 K temperature regimes. This hump can be due to the presence of bipolar effect, where both majority (holes) and minority (electrons) carriers contribute to $\kappa_{total}$ [46, 47]. Interestingly, above 650 K, $\kappa_{total}$ varies as function of $1/T$, reaching the minimum value of ~ 2.79 W/m-K near 950 K.

Heat transport mechanism in any material is primarily governed by the movement of charge carriers ($\kappa_e$) and through $\kappa_L$. However, in present case, bipolar effect ($\kappa_{bp}$) also contributes. Hence, $\kappa_{total}$ is defined as

$$\kappa_{total} = \kappa_e + \kappa_L + \kappa_{bp} \dots\dots\dots\dots\dots \quad (5)$$

Each of these components is subtracted from $\kappa_{total}$ to get the individual contributions to the thermal conductivity. The electronic contribution from $\kappa_{total}$ is extracted using:

$$\kappa_e = \frac{LT}{\rho}\ldots\ldots (6)$$

where $L$ is the Lorentz number. As ZrIrSb is a non-degenerate semiconductor, hence, variable values of $L$ are used (as shown in left inset of Fig. 4 (a)), and is defined as [48]

$$L = \left[1.5 + exp^{-\left[\frac{|S|}{116}\right]}\right]\ldots\ldots\ldots\ldots (7)$$

The temperature dependence of $\kappa_e$ is presented in the right inset of Fig. 4 (a). The value of $\kappa_e$ at 300 K is estimated to be around 0.04 W/m-K, which also increases with temperature, reaching a maximum of ~ 0.35 W/m-K near 950 K. As noted in section 3.3.2, above 750 K, $\rho$ increases with increment in temperature, implying more contribution of charge carriers to $\kappa_{total}$ at higher temperatures. In present case, there is ~ 11% contribution of charge carriers to $\kappa_{total}$ at 950 K. To estimate the contribution of $\kappa_{bp}$ to $\kappa_{total}$, method proposed by Kitagawa *et al.* is used [49]. At temperature above Debye temperature($\theta_D$), which is ~ 297 K for ZrIrSb, the Umklapp phonon scattering mechanism is expected to be dominant, where $\kappa_L$ obey the following relation [48, 50]:

$$\kappa_L = 3.5\, \frac{k_B}{h}\, \frac{MV^{1/3}\theta_D^3}{\gamma^2 T}\ldots\ldots\ldots\ldots (8)$$

where $M$ is the average mass per atom, $V$ is the average atomic volume, and $\gamma$ is the Gruneisen parameter. From the above equation, it can be said that $\kappa_L$ is proportional to $T^{-1}$. Therefore, $\kappa_{bp}$ can be estimated by subtracting the linear term from $\kappa_{total} - \kappa_e$ at higher temperatures. This method has been previously used in studying $\kappa_{total}$ of $CoSb_{3-x}Te_x$, $CoSb_3$ systems [51, 52]. The $\kappa_{total} - \kappa_e$ follow the linear behaviour of $\kappa_L \sim T^{-1}$ quite well upto 500 K (as shown in inset of Fig. 4 (b)) which confirms the dominance of Umklapp phonon scattering mechanism. It shows deviation from 500 K, and this temperature can be referred as bipolar excitation temperature. At high temperature, $\kappa_L$ is estimated by extrapolating the linear relationship of the $\kappa_L \sim T^{-1}$. The resultant curve is shown in Fig. 4 (b). The contribution of $\kappa_e$, $\kappa_{bp}$, and $\kappa_L$ to the $\kappa_{total}$ are estimated as 11.36%, 2.38% and 86.26% near 950 K, respectively.

Temperature dependent $\kappa_L$ of ZrIrSb is also calculated using anharmonic first-principles phonon calculations. In the calculation, only three phonon processes are considered due to presence of negligible separation energy gap between optical and acoustic phonon branches (discussed in details in the next paragraph). The $\kappa_L$ is estimated under SMRT approximation by solving LBTE method [33] and is defined as

$$\kappa_L = \frac{1}{NV} \sum_\lambda C_\lambda \, v_\lambda \otimes v_\lambda \, \tau_\lambda^{SMRT} \quad \ldots\ldots\ldots\ldots (9)$$

Here, $N$ is the number of unit cells in the crystal, $V_0$ is the volume of the unit cell, $c_\lambda$ is the mode-dependent specific heat, $v_\lambda$ is the group velocity of the phonon mode and $\tau_\lambda^{SMRT}$ is the single mode relaxation time of phonon mode $\lambda$. Here, $\lambda$ represents a phonon mode with a wavevector $q$, branch index $j$ and $\tau_\lambda^{SMRT}$ is considered as the lifetime of the phonon mode $\tau_\lambda$. The calculated $\kappa_L$ is presented in Fig. 4 (b). It can be observed that the calculated $\kappa_L$ decreases with the increase in temperature, which is in analogy with our experimental observations. In the whole temperature range, the calculated value of $\kappa_L$ deviates from observed experimental data. The separation between the curves reduces towards lower temperature. This discrepancy can be due to presence of the defects or anti-site disorder in ZrIrSb.

As noted above, the transport of heat in ZrIrSb is mainly through the lattice via phonons. Thus, investigation of phonon properties is essential to understand the $\kappa_L$ behaviour. Therefore, phonon dispersion along with its electronic structure is calculated using the computational method. This will help us to discern the contribution from the various phonon modes to the $\kappa_L$. The obtained phonon dispersion spectrum along high symmetric directions is presented in Fig. 4 (c). From the figure, it is inferred that, all the phonon frequencies are positive, which corroborates that the structural model of ZrIrSb is dynamically stable. For a primitive unit cell containing three atoms, there are three acoustic phonon modes (one longitudinal (LA) + two transverse (TA)) and six optical phonon modes (two longitudinal (LO) + four transverse (TO)) corresponding to those atoms. The phonon modes are classified into three categories: 0-3.5 THz (acoustic modes; Black solid lines), 3.5- 5.25 THz (low frequency optical modes; Red solid lines) and 5.5- 7 THz (high frequency optical modes; Blue solid lines). From Fig. 4 (c), it can be inferred that the separation gap between acoustic and low frequency optical modes is very small ~ 0.64 meV, whereas there is presence of large gap between low frequency and high frequency optical modes. Due to presence of lower energy gap between acoustic and low frequency optical modes, it is expected that low frequency optical modes will contribute to $\kappa_L$. The low frequency optical modes lead to an increment in Umklapp processes which tends to reduce $\kappa_L$.

Additionally, to extract the contribution of different atoms to the acoustic and optical phonon modes, phonon total DOS (P-TDOS) along with the partial DOS (P-PDOS) are calculated. P-TDOS per unit cell and P-PDOS per atom are shown in Fig. 4 (d). In analogy to Fig. 4 (c), in P-TDOS plot, a separation gap between higher frequency and lower frequency optical phonon modes can be noted. From the plots of P-PDOS, it can be noted that higher frequency optical phonon modes (~ 5.5-7 THz) are mainly due to lighter Zr atoms. The considerable number of phonon states in the mid frequency range (~ 3.5- 5.25 THz) are due to Sb and Ir atoms. However, the main contribution to lower frequency acoustic phonon modes (~ 0- 3.5 THz) comes from heavier Ir atoms in the unit cell.

Moreover, from our experimental results it is noted that the value of $\kappa_L$ near room temperature is quite less [38]. This signifies that the presence of heavier elements has some significant impact on the phonon dispersion of ZrIrSb. The most remarkable difference is noted in the group velocities of acoustic phonon modes. The group velocities of phonon modes are estimated close to Γ point. The estimated values are ~ 2.6 km/s/2.6 km/s/5.2 km/s (TA/TA/LA). Additionally, flattening of the high frequency optical phonon branches in the vicinity of Γ point has been noted in Figure 4 (c). It signifies that the group velocities (TA/LA) of high frequency optical phonon modes are negligible. Therefore, the group velocities of acoustic phonon modes can be approximated as sound velocity of the alloy. The calculated values are comparable with the experimentally obtained longitudinal ($v_L$) and transverse ($v_t$) sound velocities (given in table 5). However, as noted from table 5, the values of $v_L$ and $v_t$ are smaller than that reported for TiCoSb, ZrCoSb and HfCoSb. The reduced sound velocity is due to the fact that, in ZrIrSb, in addition to acoustic phonon modes, the low frequency optical phonon modes also play a significant role. In addition to this, a distinct variation is seen in the frequency of optical phonon modes. As noted from Fig. 4(c), the lowest frequency optical modes lie near 5 MHz, whereas in TiCoSb, ZrCoSb and HfCoSb, it lies near 6.5 MHz, 5.9 MHz and 5.2 MHz, respectively [53]. Due to the low lying optical phonon modes, it is expected that there will be strong coupling between optical and acoustic phonon modes, which is absent in the latter alloys. This leads to an increment in Umklapp processes of phonon scattering, which is observed due to the presence of large number of optical phonon excitations near the Brillouin zone boundary. Hence, it can be concluded that the lower $\kappa_L$ noted in this alloy is due to the reduced velocity of phonons and enhancement in Umklapp process of phonon scattering.

Ideally, in semiconductors or insulators, the acoustic modes play a significant role in the conduction of heat. However, as discussed above in the present case, the interaction of optical phonon modes with acoustic vibrations cannot be neglected. Hence, the contributions of different phonon modes to $\kappa_L$, is analysed through accumulated lattice thermal conductivity ($\kappa_C$). The accumulated thermal conductivity ($\kappa_C$) is defined as [54]

$$\kappa_C(\omega) = \int_0^\omega \kappa_L \, \delta(\omega' - \omega_\lambda) d\omega' \dots \dots \dots (10)$$

where $\kappa_L$ is given by eqn. 9. The predicted normalized $\kappa_C$ with respect to $\kappa_L$ as function of frequency is presented in Fig. 5. At 300 K, the contribution of acoustic phonon modes (0-3.5 THz) and lower optical phonon modes (3.5 THz- 5.25 THz) is found to be ~ 50% and 45% respectively. However, higher optical phonon modes contribute only 5% to $\kappa_C$. Hence, approximately, 95% of the calculated $\kappa_L$ is from delocalized low frequency phonon modes while higher optical modes contribute insignificantly to the $\kappa_L$. This indicates that the phonon modes arising due to heavier elements dominate the heat transport mechanism. It is estimated to be ~ 80% at 300 K. Interestingly, as the temperature increases from 300 to 500 K, $\kappa_C/\kappa_L$ reduces due to increase in phonon-phonon interactions and higher scattering rates, whereas contribution from Ir and Sb phonon modes increases. Similar variation is noted from 500-700 K and 700-900 K temperature range.

Additionally, the $\kappa_L$ is also sensitive to the magnitude and frequency dependence of relaxation time ($\tau$). Hence, phonon lifetime as a function of phonon frequency at 300, 500, 700 and 1000 K is represented in Fig. 6. The phonon lifetime is calculated from anharmonic three phonon-phonon scattering mechanisms or Umklapp processes. A sparse distribution of phonon modes can be observed in acoustic and lower optical modes regime. At 300 K, the longest lifetime ($\tau_{max}$) ~ 32 ps is noted for phonons belonging to acoustic modes. $\tau$ is found to decrease with increment in frequency due to enhancement in phonon scattering events. Similar distribution of modes and trend in $\tau$ is noted in lower optical modes regime. It indicates that phonons from both acoustic and optical modes contribute to $\kappa_L$ which is in analogy with previous observations. However, above 5.5 THz, the rate of decrement of $\tau$ is reduced. This behaviour is observed because the phonon number density increases (yellow/orange region) leading to higher scattering rates. Increment in scattering rate subsequently leads to the phonon annihilation and ultimately decreases the lifetime of decay. Further, as the temperature increases from 300 to 500 K, $\tau_{max}$ decreases which indicates an enhanced phonon-phonon scattering rates at higher temperature. Hence a reduced $\kappa_L$ is noted. Similar behaviour is noted in 500-700 K and 700-1000 K temperature regime. Therefore, from our investigations, it can be said that there is existence of coupling between the acoustic and lower lying optical phonon modes originating due to Ir and Sb atoms, which plays a critical role in the thermal transport mechanism of ZrIrSb. This behaviour is quite different from the traditional understanding of thermal transport. In our case, a lower $\kappa_L$ is expected due to contribution from both modes but owing to large distribution of phonon modes in this regime (0-5.25 THz), a large $\kappa_L$ is observed. Additionally, the decrement in $\kappa_L$ with increment in temperature can be attributed to decrement in τ of acoustic and lower optical phonon modes which is an outcome of enhanced phonon-phonon scattering rate.

Finally, in order to evaluate the thermoelectric efficiency of the material, the ZT is estimated using the following equation [48]

$$ZT = \frac{S^2 T}{\rho \kappa_{total}} \ldots\ldots\ldots\ldots (11)$$

The ZT obtained experimentally for ZrIrSb is shown as function of temperature in Figure 7 (a). ZT value for ZrIrSb is ~ 0.032 at 950 K which is larger than those reported for similar alloys like TiCoSb, ZrCoSb and HfCoSb [37, 38]. The large value for ZrIrSb can be due to the lower value of $\rho$ and $\kappa_L$. To enable this alloy to be used for thermoelectric conversion, an efficient method is to further decrease its $\kappa_L$. Various strategies like nano-structuring, construction of heterojunctions, introducing mass or strain effect has been employed by researchers [55, 56]. Among these, we have further discussed the possibility of employing nano-structuring as an effective tool to decrease the $\kappa_L$. In this strategy, the particle size is reduced which in turn result in the decrement of the phonon mean free path (MFP). Hence, to explore this possibility, the size effect on the $\kappa_L$ is evaluated through the summation of contributions from all phonon modes in ZrIrSb. The ratio $\kappa_C/\kappa_L$ plotted as a function of MFP is shown in Fig. 7 (b). At 300 K, it is noted that 10 nm size crystals can reduce the $\kappa_L$ by ~87% as compared to the bulk, which in turn may increase the ZT substantially. Similar improvements are achievable at 500 K, 700 K and 900 K operating temperatures. Thus, it would be interesting to explore the thermoelectric properties of similar alloys through nano-structuring.

### 4. Conclusions

In conclusion, the combined experimental and theoretical investigation on ZrIrSb reveals that this alloy is a *p*-type semiconductor. It has been observed that the magnitude of electronic parameters ($\rho$ and *S*) is adversely affected by the anti-site disorder between heavier elements (Ir/Sb) and vacant sites. Interestingly, in this alloy, $\kappa_L$ is dominated by Umklapp processes originating due to the coupling between acoustic and low frequency optical phonon modes, resulting in a reduction in heat conduction. Additionally, the obtained ZT is among the highest as noted for analogues alloys. Our investigation may provide a pathway for researchers to explore the thermoelectric properties of heavy constituents HH alloys through nano-structuring for potential thermoelectric applications.

**Acknowledgements**

KM acknowledges financial support from DST-SERB project CRG/2020/000073.

**Tables:**

**Table 1.** Obtained structural parameters from Rietveld refinement of ZrIrSb

| Experimental lattice parameter (Å) | 6.292±0.001 | | | |
|---|---|---|---|---|
| **Theoretical lattice parameter (Å) (WEIN2K code)** | 6.3672 Å | | | |
| **Theoretical lattice parameter (Å) (VASP code)** | 6.3520 Å | | | |
| **Volume (Å$^3$)** | 249.45 | | | |
| **Wyckoff positions** | X | Y | Z | position |
| **Zr** | 0 | 0 | 0 | **4a** |
| **Ir** | 0.25 | 0.25 | 0.25 | **4c** |
| **Sb** | 0.50 | 0.50 | 0.50 | **4b** |
| **Vacant sites** | 0.75 | 0.75 | 0.75 | **4d** |

**Table 2.** The effective mass of holes (at Γ point) and electrons (at X point) along the high symmetric direction

| High symmetry point | Effective mass ($m^*/m_e$) | | | |
|---|---|---|---|---|
| | Valence Band | | | Conduction Band |
| | **Band 1** | **Band 2** | **Band 3** | **Band 4** |
| **K-Γ-K** | 0.16(0) | 0.55(0) | 4.32(0) | |
| **W-Γ-W** | 0.18(4) | 0.57(4) | 1.62(2) | |
| **X-Γ-X** | 0.21(7) | 0.6(0) | 0.59(1) | |
| **L-Γ-L** | 0.15(0) | 1.42(4) | 1.40(9) | |
| | | | | |
| **K-X-K** | | | | 0.36 |
| **W-X-W** | | | | 0.34 |
| **Γ-X-Γ** | | | | 0.60 |
| **L-X-L** | | | | 0.41 |

**Table 3.** Comparison between $\rho_{300\ K}$ of different thermoelectric materials

| Alloy | Value (mΩ-cm) | |
|---|---|---|
| **ZrCoSb** | 80 | Ref.38 |
| **HfCoSb** | 20 | Ref.37 |
| **TiCoSb** | 100 | Ref.17 |
| **Bi$_2$Se$_3$** | 0.2 | Ref.41 |
| **Cu$_{0.1}$Bi$_2$Se$_3$** | 6.8 | Ref.41 |
| **Sb$_2$Te$_3$** | 2.5 | Ref.42 |
| **Bi$_{0.9}$Sb$_{0.1}$** | 0.01 | Ref.43 |

**Table 4.** Comparison with thermal conductivity of XCoSb (X= Ti, Zr and Hf) alloys with ZrIrSb at 300 K

| Alloy | Value of $\kappa_{total}$ (W/m-K) | |
|---|---|---|
| TiCoSb | 24 | Ref. 37 |
| ZrCoSb | 22 | Ref. 37 |
| HfCoSb | 12 | Ref. 37 |
| ZrIrSb | 9.5 | This work |

**Table 5.** Longitudinal and transverse sound velocities of ZrIrSb and XCoSb (X= Ti, Zr and Hf)

| Alloy | | Longitudinal (m/s) | Transverse (m/s) | |
|---|---|---|---|---|
| **ZrIrSb** | **Experimental** | 4479 | 2486 | This work |
| | **Calculated** | 5203 | 2670 | This work |
| **TiCoSb** | **Experimental** | 5691 | 3230 | Ref. 38 |
| **ZrCoSb** | **Experimental** | 5488 | 3144 | Ref. 38 |
| **HfCoSb** | **Experimental** | 4703 | 2709 | Ref. 38 |

**Figures**

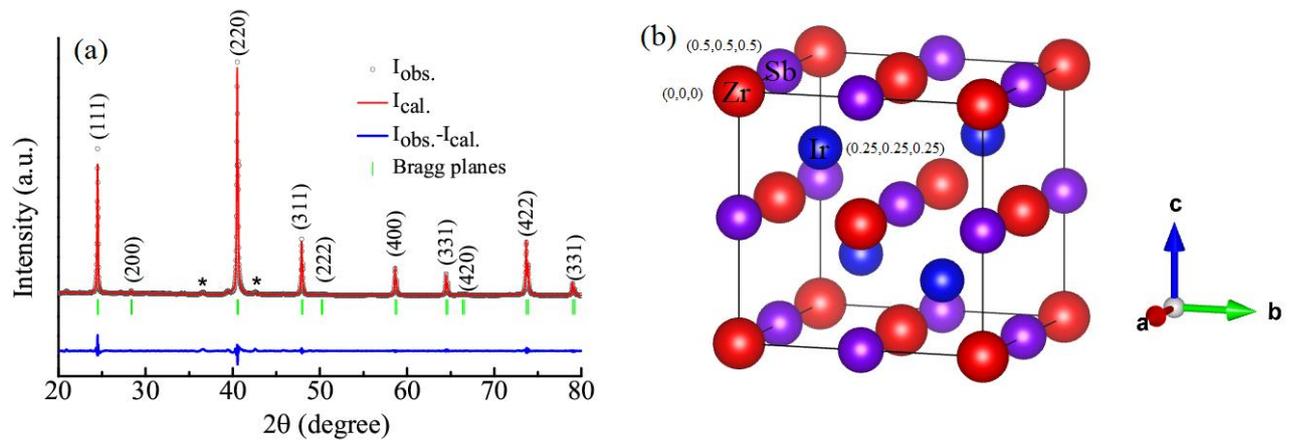

**Figure 1.** (a) Rietveld refinement analysis of room temperature XRD pattern (Asterisks indicate unidentified peaks). (b) Crystal structure of ZrIrSb

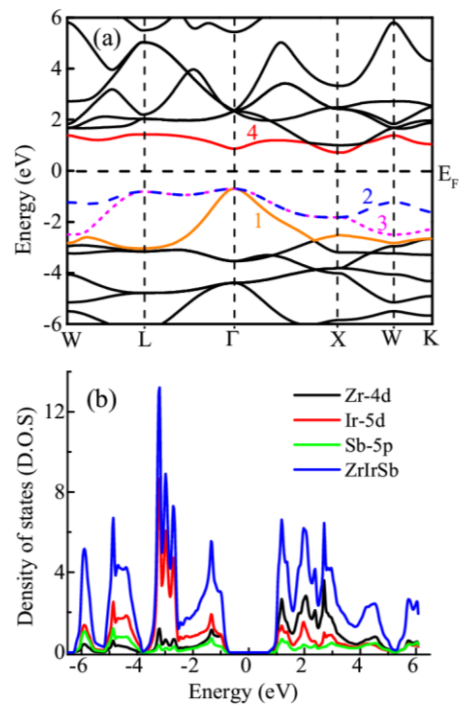

**Figure 2.** (a) Electronic dispersion of ZrIrSb along high symmetric *k*-points (b) The TDOS (in states/eV/f.u) and PDOS in (states/eV/atom) of ZrIrSb

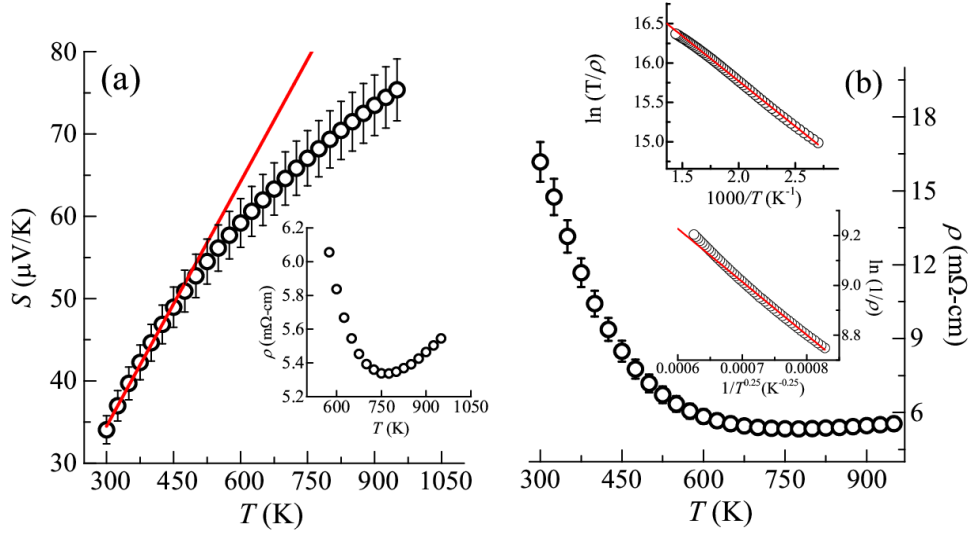

**Figure 3.** (a) Temperature dependent behaviour of $S$ in the temperature regime 300-950 K of ZrIrSb; Red solid line represents linear trend of $S$ Inset: Inset: $\rho$ vs $T$ plot in the temperature regime 550- 950 K. (b) Temperature response of $\rho$ in the temperature regime 300-950 K for ZrIrSb; Upper Inset: $\ln(T/\rho)$ vs $1000/T$ (K$^{-1}$) plot; Red solid line represents linear fitting. Lower Inset: $\ln(1/\rho)$ vs $1/T^{0.25}$ plot; Red solid line represents linear fitting.

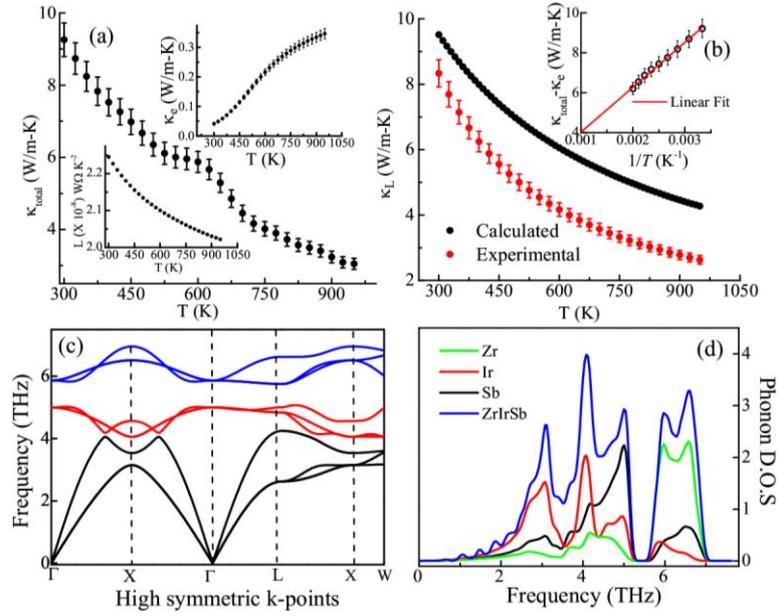

**Figure 4.** (a) $\kappa_{total}$ as function of temperature; Left inset: temperature dependent behaviour of $L$. Right inset: Temperature dependent behaviour of $\kappa_e$ in the temperature range 300-950 K. (b) Temperature response of calculated and experimental $\kappa_L$ of ZrIrSb in the temperature range 300-950 K Inset: $\kappa_{total}$ - $\kappa_e$ as a function of temperature. (c) Phonon dispersion curve of ZrIrSb along high symmetric $k$-points (d) Total (in states/eV/f.u) and P-PDOS (in states/eV/atom) of ZrIrSb.

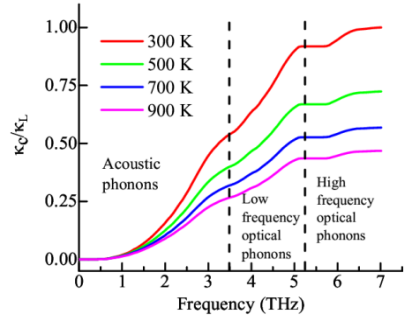

**Figure 5.** The normalized $\kappa_C(\omega)$ as a function of frequency (in THz) at four temperatures.

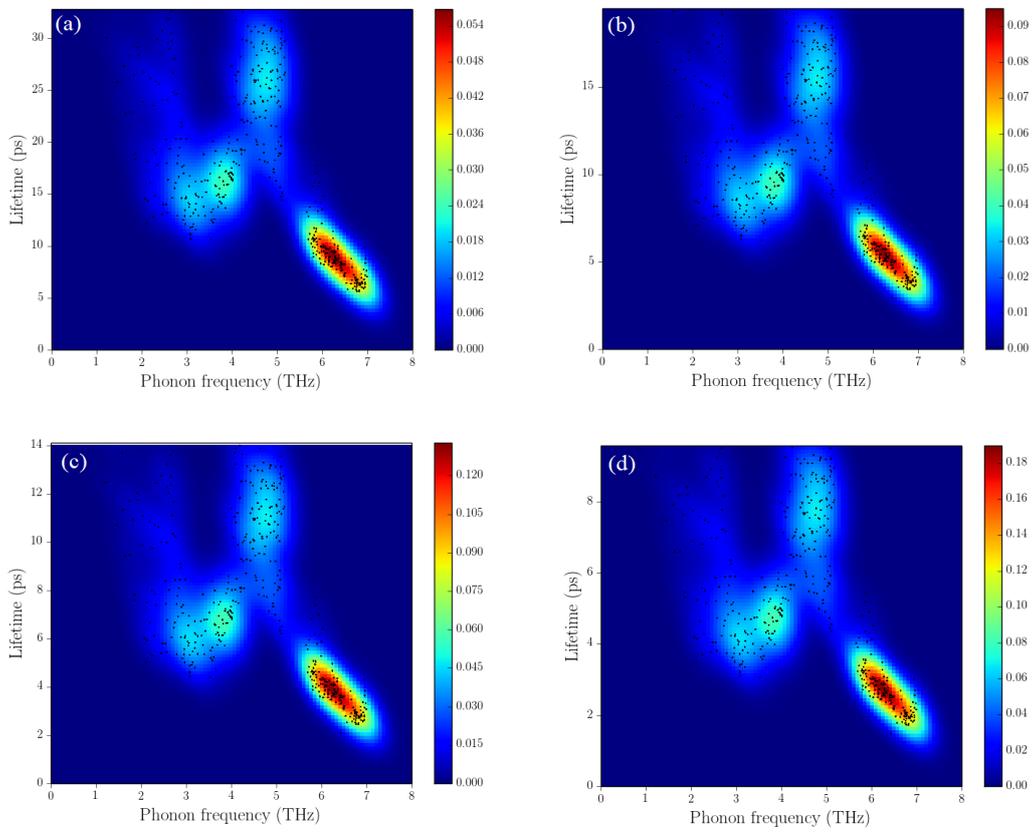

**Figure 6.** Phonon lifetime as a function of phonon frequency at (a) 300 K (b) 500 K (c) 700 K and (d) 1000 K. The phonon modes are displayed in black while the colored background represents the density of the phonon modes in the corresponding region.

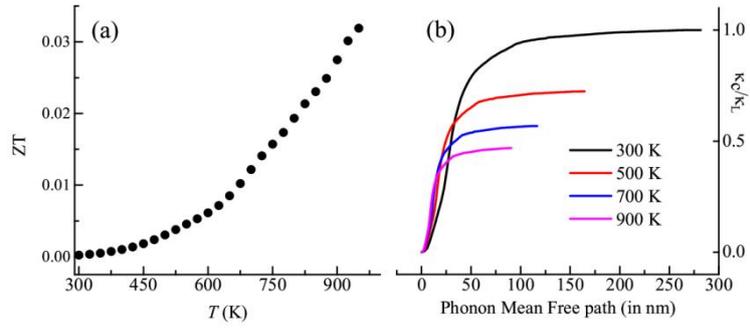

**Figure 7.** (a) ZT as function of temperature in the temperature regime 300-950 K (b) The normalized $\kappa_C(\omega)$ as a function of phonon MFP (in nm) at four temperatures.